\documentclass[english,amssymb,floatfix]{revtex4}
\usepackage[T1]{fontenc}
\usepackage[latin9]{inputenc}
\usepackage{amsmath}
\usepackage{amssymb}
\usepackage{graphicx}
\usepackage{esint}

\makeatletter

\providecommand{\tabularnewline}{\\}

\@ifundefined{textcolor}{}
{%
 \definecolor{BLACK}{gray}{0}
 \definecolor{WHITE}{gray}{1}
 \definecolor{RED}{rgb}{1,0,0}
 \definecolor{GREEN}{rgb}{0,1,0}
 \definecolor{BLUE}{rgb}{0,0,1}
 \definecolor{CYAN}{cmyk}{1,0,0,0}
 \definecolor{MAGENTA}{cmyk}{0,1,0,0}
 \definecolor{YELLOW}{cmyk}{0,0,1,0}
}

\usepackage{amsfonts}\usepackage{bbm}\usepackage{epsfig}\usepackage{graphics}
\usepackage{ifpdf}\ifpdf
\DeclareGraphicsExtensions{.pdf,.png}\usepackage{pgf}\usepackage{tikz}\usepackage{epstopdf}

\else
\DeclareGraphicsExtensions{.eps,.png}
\fi

\graphicspath{{./}}

\newcommand{\lsim}{\,\lower2truept\hbox{${<\atop\hbox{\raise4truept\hbox{$\sim$}}}$}\,}\newcommand{\gsim}{\,\lower2truept\hbox{${>\atop\hbox{\raise4truept\hbox{$\sim$}}}$}\,}

\newcommand{\be}{\begin{equation}}\newcommand{\ee}{\end{equation}}\newcommand{\bea}{\begin{eqnarray}}\newcommand{\eea}{\end{eqnarray}}\newcommand{\beann}{\begin{eqnarray*}}\newcommand{\eeann}{\end{eqnarray*}}

\newcommand{\dd}{{\rm d}}

\newcommand{\eprint}[1]{\url{arXiv:#1}}

\makeatother

\usepackage{babel}

\makeatother

\usepackage{babel}

\makeatother

\usepackage{babel}

\makeatother

\usepackage{babel}

\makeatother

\usepackage{babel}
\begin{document}

\title{Testing coupled dark energy with next-generation large-scale observations}

\author{Luca Amendola$^{1}$, Valeria Pettorino$^{2}$, Claudia Quercellini$^{3}$,
Adrian Vollmer$^{1}$}

\affiliation{$^{1}$ Institut fuer Theoretische Physik, Universitaet Heidelberg,
Philosophenweg 16, D-69120 Heidelberg, Germany. \\
 $^{2}$ SISSA, Via Bonomea 265, 34136 Trieste, Italy, \\
 $^{3}$ University of Rome Tor Vergata, Via della Ricerca Scientifica,
1 - I-00133 Roma (Italy) }
\begin{abstract}
Coupling dark energy to dark matter provides one of the simplest way
to effectively modify gra\-vity at large scales without strong constraints
from local (i.e. solar system) observations. Models of coupled dark
energy have been studied several times in the past and are already
significantly constrained by cosmic microwave background experiments.
In this paper we estimate the constraints that future large-scale
observations will be able to put on the coupling and in general on
all the parameters of the model. We combine cosmic microwave background,
tomographic weak lensing, redshift distortions and power spectrum
probes. We show that next-generation observations can improve the
current constraint on the coupling to dark matter by two orders of
magnitude; this constraint is complementary to the current solar-system bounds on a coupling
 to baryons.
\end{abstract}

\date{\today}

\maketitle

\section{Introduction}

Several observational campaigns will be launched in the next few years
to advance our knowledge of the matter distribution from redshifts
of order $z\le0.5$ (the depth of the SDSS \cite{sdss} and 2dF \cite{2df}
surveys) to $z\le2-3$ (the range of satellite missions like Euclid
\cite{euclidredbook} and of ground-based surveys like LSST \cite{lsst},
DES \cite{des}, etc). These surveys will map hundreds of millions
of galaxy redshifts and billions of galaxy images distributed across
almost the full accessible sky. The information contained in these
data will occupy the cosmologists for years to come and it is very
useful already at this stage to foresee what these data can tell us
about the currently investigated cosmological models. Several works
have been devoted to forecasting future constraints with future wide
surveys and a small selection of the recent ones can be found in Ref.
\cite{Wang2010,martinelli,diporto10,DeBernardis:2011iw,carbone_etal_2011}.

The main science drivers for most of these large-scale enterprises
is the quest for dark energy (DE). Even if we get very accurate knowledge
of the cosmic background expansion and therefore of the dark energy
equation of state, this will not be enough to distinguish among competing
models. In fact, in the models in which dark energy clusters or gravity
is modified (see e.g. \cite{Amendola2010}), the linear growth of
clustering is in general not completely fixed by background observables.
Mapping the linear clustering has been identified as the main probe
of modification of gravity or, more in general, of non-standard dark
energy models.

The main probes of matter clustering are the weak lensing and the
galaxy power spectrum $P(k)$ and indeed most large scale surveys
aim at mapping one or both of these quantities. For instance Euclid,
probably the most ambitious project to date, sets as goal the mapping
of half the sky both in imaging and in broad-band spectroscopy, with
an average redshift of order unity and a maximal useful redshift around
2 or 3. Combining weak lensing and $P(k)$ gives the opportunity to
break several degeneracies among the cosmological parameters and to
directly measure the growth function.

Both WL and $P(k)$ probe the universe at relatively small redshifts.
A much longer lever arm is obtained by combining them with the cosmic
microwave background, in particular employing the specifics of the
Planck mission. In this way we can tighten sensibly the constraints
on all the cosmological parameters.

In this paper we focus on one particular class of dark energy models
that includes typical features of modified gravity models, the coupled
dark energy model (\cite{Wetterich1994,Amendola1999,Pettorino:2008ez},
see e.g. \cite{Amendola2010} for a review). The idea behind this
model is very simple: when dark energy is promoted to a scalar field
$\phi$, as in the vast majority of DE models and also as in the prototypical
accelerated expansion, inflation, one naturally wonders whether this
field could be mediating a force among matter particles. This is indeed
the same phenomenology as in the Brans-Dicke model, except that now
we require this field also to drive acceleration and therefore to
be dominant and to carry a potential. In this paper we model the $\phi$
potential as an inverse power-law $V\sim\phi^{-\alpha}$, as in the
Ratra-Peebles scenario \cite{Peebles2003}. This one-parametric family
is simple but gives us enough freedom to explore the dependence of
the results on the potential slope and therefore on the equation of
state. The phenomenology of this interaction can be immediately grasped
by writing down the energy-momentum conservation equations: 
\begin{align}
T_{(m)\nu:\mu}^{\mu} & =\beta T_{(m)}\phi_{;\nu}\\
T_{(\phi)\nu:\mu}^{\mu} & =-\beta T_{(m)}\phi_{;\nu}
\end{align}
 where the subscripts $m,\phi$ denote matter and dark energy, respectively
and where $\beta$ is the coupling constant. While $\beta$ could
also be a coupling function $\beta(\phi)$ (see e.g. \cite{bonometto2007,Baldi:2010vv,beyer2011}),
in this paper we restrict ourselves for simplicity to a constant value.
A coupled particle of mass $m$ will respond to this interaction by
following in the Newtonian regime the scalar-gravitational potential
\begin{equation}
U=-\frac{Gm}{r}(1+2\beta^{2}e^{-m_{\phi}r})
\end{equation}
 where $m_{\phi}$ depends on the field potential. For a dark energy
field one expects the interaction scale $1/m_{\phi}$ to be of astrophysical
size. The parameter $\beta^{2}$ quantifies therefore the deviation
from Newtonian dynamics. Any indication that $\beta\neq0$ would of
course show a fundamental modification of nature's law. The main goal
of this paper is to forecast the future constraints on $\beta$ and
to see how all the other cosmological parameters depend on it.

\begin{figure}
\includegraphics[width=10cm]{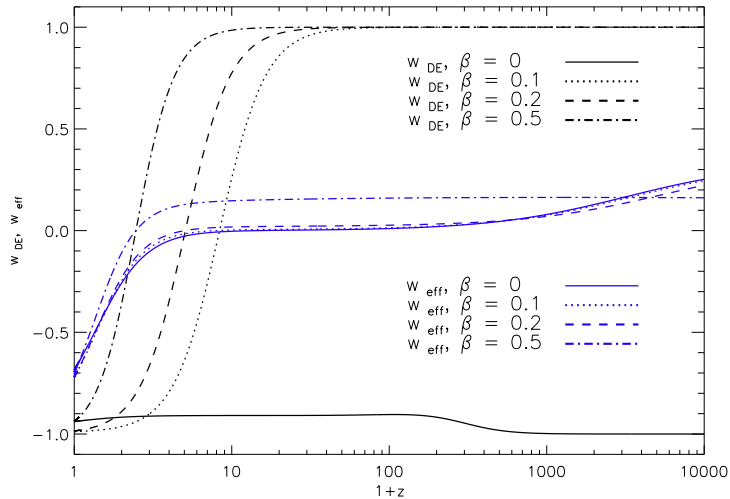} 

\caption{\label{fig:Evolution-of-}Evolution of $w_{\mathrm{eff}}$ for various
values of $\beta$.}
\end{figure}

Since gravity now is supplemented by an additional scalar mediator,
the resulting scalar-tensor force deviates from Einstein's (and from
Newton's) one and this gives rise to a number of observable differences
with respect to the uncoupled models, both at the background and at
the perturbation level. The expansion is generally modified when $\beta$
is non-zero. In particular, it has been shown in \cite{amendola_etal_2003}
that for most potentials DE is not negligible in the past, contrary
to the cosmological constant case and to models that do not drastically
deviate from it. Adopting the Ratra-Peebles potential one finds that
the evolution of the scalar field during the matter domination is
controlled by the coupling $\beta$ and the density fraction $\Omega_{\phi}$
is proportional to $\beta^{2}$. As a consequence, the expansion rate
$H$, given in terms of the effective equation of state $w_{\mathrm{eff}}$
by 
\begin{equation}
\frac{H'}{H}=-\frac{3}{2}(1+w_{\mathrm{eff}})
\end{equation}
 is modified. We define here $w_{\mathrm{eff}}\equiv p_{t}/\rho_{t}$,
where $p_{t}$ and $\rho_{t}$ are the total pressure and energy density
contributions respectively. The value of $w_{eff}$ decreases from
$1/3$ in radiation dominated era to 
\begin{equation}
w_{\mathrm{eff}}\approx\frac{2}{3}\beta^{2}
\end{equation}
 during matter dominated epoch, where the dominant contribution to
$p_{t}$ is given by the scalar field. More recently, dark energy
dominates and $w_{DE}$ tends to $-1$: the higher the coupling, the
fastest is the transition between matter and dark energy dominated
era. This behavior is illustrated in Fig.(\ref{fig:Evolution-of-})
Accordingly, properties as the age of the Universe and the distance
to last scattering (and therefore the position of the acoustic peaks)
do also depend on $\beta$ and this allows to put stronger constraints
on it. In fact the best constraints on $\beta$ so far come from the
CMB data of WMAP, $|\beta|\le0.15$ \cite{amendola_quercellini_2003,Bean:2008ac}.
We stress that here we consider massless neutrinos but it is important
to notice that if neutrinos are massive higher values of the coupling
($\beta\le0.2$) can be allowed \citep{LaVacca:2009yp}. At the perturbation
level, the extra pull induced by the scalar force adds to gravity
and speeds up the perturbation growth, resulting in further possible
constraints.

Any modification of gravity has to pass the strong bounds from local
experiments, both in laboratory and in the solar system or other astrophysical
sources (see e.g. \cite{2003RvMP...75..403U,2007PhRvL..98b1101K,Nakamura:2010zzi}).
This can be achieved in several ways by the so-called chameleon \cite{chameleon,Hui:2009kc}
or Vainshtein mechanisms \cite{vainshtein} but here we adopt a simpler
way out: we assume that baryons are uncoupled. In this case, all local
experiments are bypassed and cosmology becomes the only way to observe
the coupling.

\section{Coupled dark energy}

\label{cde}

We consider coupled dark energy cosmologies as described by the lagrangian:
\begin{equation}
{\cal L}=-\frac{1}{2}\partial^{\mu}\phi\partial_{\mu}\phi-U(\phi)-m(\phi)\bar{\psi}\psi+{\cal L}_{{\rm kin}}[\psi]\,,\label{L_phi}
\end{equation}
 in which the mass of matter fields $\psi$ coupled to the DE is a
function of the scalar field $\phi$. In the following we will consider
the case in which DE is only coupled to cold dark matter (CDM, hereafter
denoted with a subscript $c$). The choice $m(\phi)$ specifies the
coupling and as a consequence the source term $Q_{(\phi)\mu}$ via
the expression: 
\begin{equation}
Q_{(\phi)\mu}=\frac{\partial\ln{m(\phi)}}{\partial\phi}\rho_{c}\,\partial_{\mu}\phi.
\end{equation}
 Due to the constraint of conservation of the total energy-momentum
tensor, if no other species is involved in the coupling, $Q_{(c)\mu}=-Q_{(\phi)\mu}$.

The conservation equations for the energy densities of each coupled
species are: 
\begin{eqnarray}
\rho_{\phi}' & = & -3{\cal H}\rho_{\phi}(1+w_{\phi})-Q_{(\phi)0}\,\,\,\,,\label{rho_conserv_eq_phi}\\
\rho_{c}' & = & -3{\cal H}\rho_{c}+Q_{(\phi)0}\nonumber 
\end{eqnarray}
 (where primes denote differentiation with respect to conformal time
and ${\cal H}=aH$ is the conformal Hubble function) plus the standard
conservation equation for baryons. Here we have treated each component
as a fluid with ${T^{\nu}}_{(\alpha)\mu}=(\rho_{\alpha}+p_{\alpha})u_{\mu}u^{\nu}+p_{\alpha}\delta_{\mu}^{\nu}$,
where $u_{\mu}=(-a,0,0,0)$ is the fluid 4-velocity and $w_{\alpha}\equiv p_{\alpha}/\rho_{\alpha}$
is the equation of state. The class of models considered here corresponds
to the choice: 
\begin{equation}
m(\phi)=m_{0}e^{-\beta\frac{\phi}{M}}\,,\label{coupling_const}
\end{equation}
 with the coupling term equal to 
\begin{equation}
Q_{(\phi)0}=-\frac{\beta}{M}\rho_{c}\phi'\,.
\end{equation}
 Equivalently, the scalar field evolves according to the Klein-Gordon
equation: 
\begin{equation}
\phi''+2{\cal H}\phi'+a^{2}\frac{dV}{d\phi}=a^{2}\beta\rho_{c}\,\,.\label{kg}
\end{equation}
 Throughout this paper we choose an inverse power law potential defined
as: 
\begin{equation}
V=V_{0}\phi^{-\alpha}
\end{equation}
 with $\alpha$ and $V_{0}$ constants. When the coupling is absent,
this potential leads to a well-known transient tracking solution \cite{Wetterich1994,Amendola1999,Copeland:1997et}
in which 
\begin{equation}
w_{DE}\approx-\frac{2}{\alpha+2}
\end{equation}
 which sets up just before the final dark energy domination; after
this regime, asymptotically $w_{DE}\to-1$. When $\beta\not=0$ but
small, this tracking regime is also approximately true but squeezed
between the modified matter era with $w_{\mathrm{eff}}\approx2\beta^{2}/3$
and the final phase with $w_{DE}\to-1$ and therefore barely visible.
Current supernovae Ia constraints impose an upper limit $\alpha\le0.2$.

Various choices of couplings have been investigated in literature
\cite{Wetterich1994,Amendola1999,Mangano_etal_2003,Amendola2004,Koivisto:2005nr,Guo:2007zk,Pettorino:2008ez,quercellini_etal_2008,Quartin:2008px,Valiviita:2009nu}. Analysis of the models and constraint on these couplings have been obtained in several
ways, including spherical collapse (\cite{Wintergerst:2010ui, Mainini:2006zj} and references therein), time renormalizazion group \cite{Saracco:2009df}, $N$-body simulations \cite{Baldi:2010td,Baldi_etal_2010,Baldi:2010vv}
and effects on supernovae, CMB and cross-correlation of CMB and LSS
\cite{amendola_etal_2003,amendola_quercellini_2003,Bean:2008ac,LaVacca:2009yp,Kristiansen:2009yx,Mainini:2010ng,DeBernardis:2011iw,Xia:2009zzb,martinelli}.
Refs. \cite{martinelli,DeBernardis:2011iw} are particularly similar
in spirit to our work. However there are several differences. First,
the coupling is different since we use the coupling induced by a scalar-tensor
model. Second, we include the galaxy clustering probe (baryon acoustic
oscillations or BAO, redshift distortions and full $P(k)$ shape).
Third, we include the potential slope $\alpha$ as an additional parameter.
Fourth, we use the updated Euclid Definition phase \cite{euclidredbook}
specification (there is some change with respect to the ones in the
Assessment phase defined in the yellow book \cite{euclidyellowbook};
in particular, the total area has been reduced from 20,000 to 15,000
square degrees).

\section{Methodology}

\label{sec:methods}

In this section we present the methodology we used to derive predictions
on cosmological parameter constraining power from a combination of
upcoming and future datasets, namely galaxy power spectrum sliced
in several redshift bins, CMB angular power spectra (temperature and
polarization) and Weak Lensing power spectrum. The statistical method
chosen to do forecasts is the Fisher Matrix analysis. In order to
derive the contour confidence regions in the parameter space both
the likelihood of the data and the distribution of parameter values
is assumed to be Gaussian.

Suppose to have an observable $\mathbf{X_{obs}}=[X_{1},...,X_{N}]$
which is a function of the parameter set $\boldsymbol{\Theta}=[\theta_{1},...,\theta_{n}])$.
Then the probability of estimating unknown parameters based on known
outcomes is 
\begin{equation}
\mathcal{L}(\mathbf{X_{obs}}/\boldsymbol{\theta})\propto\exp{(\mathbf{X_{obs}}-\mathbf{X}(\boldsymbol{\theta}))^{T}\mathcal{C}^{-1}(\mathbf{X_{obs}}-\mathbf{X}(\boldsymbol{\theta}))}
\end{equation}
 where $\mathbf{X}(\boldsymbol{\theta})$ is the theoretical prediction
for $\mathbf{X_{obs}}$ as a function of $\boldsymbol{\theta}$ and
$\mathcal{C}$ is the covariance matrix of the observed components.
By assuming priors on each parameter, one is in principle able to
sample this function at reasonable accuracy level.

In the Fisher Matrix formalism the observed outcome is the mean values
of the observables $\mathbf{X}_{\mu}$ assumed as the null hypothesis.
This method allows a quick way to estimate errors on cosmological
parameters, given errors in observable quantities. The Fisher matrix
is defined as the Hessian of the log-likelihood function $\mathcal{L}$,
\begin{equation}
F_{ij}=\left<-\frac{\partial^{2}\log{\mathcal{L}(\mathbf{X}_{\mu}/\boldsymbol{\theta})}}{\partial\theta_{i}\partial\theta_{j}}\right>
\end{equation}
 such that if the parameters are assumed to be gaussianly distributed
their likelihood can be written as 
\begin{equation}
\mathcal{L}(\boldsymbol{\theta})\propto\exp{-\frac{1}{2}\sum_{ij}\theta_{_{i}}F_{ij}\theta_{j}}.
\end{equation}

By the Cramer-Rao inequality, a model parameter $\theta_{i}$ cannot
be measured to a precision better than $(F_{ii})^{-1/2}$ when all
other parameters are fixed, or a precision $((F^{-1})_{ii})^{1/2}$
when all other parameters are marginalized over. In practice, the
Fisher matrix is a good approximation to the uncertainties as long
as the likelihood can be approximated by a Gaussian, which is generally
the case near the peak of the likelihood and therefore in cases when
the parameters are measured with small errors. Conversely, if the
errors are large, then the likelihood is typically non-Gaussian, and
the constraint region is no longer elliptical but characteristically
banana-shaped. In this case, the Fisher matrix typically underestimates
the true parameter errors and degeneracies, and one should employ
the full likelihood calculation approach to error estimation.

In this work the observables $\mathbf{X}$ correspond to the galaxy
power spectrum measured by a future experiment like Euclid, the imminent
Planck CMB angular power spectra (TT, EE, BB and TE components) and
the weak lensing power spectrum as detected by a future Euclid-like
mission. The cosmological parameter set $\boldsymbol{\theta}$ comprises
the dimensionless Hubble parameter $h$, the dark matter and baryon
density parameters $\Omega_{c}$ and $\Omega_{b}$, respectively,
the spectral index of primordial perturbations $n_{s}$, the coupling
constant $\beta$ and the slope of the dark energy potential $\alpha$.
We assume a flat geometry. In the Fisher formalism, all the information
about the experiment is embedded in the covariance matrix: hence we
need to define the suitable covariance for each observable/experiment.

As a fiducial model we assume the values of parameters shown in Tab.\ref{cosmological_parameters}.
Other fiducial parameters needed for the various probes are specified
later on. Since the main cosmological effects depend on $\beta^{2}$
(for small $\beta$) we use $\beta^{2}$, rather than $\beta$, as
parameter. This confines the parameter volume to $\beta^{2}\ge0$.
Choosing the fiducial at $\beta^{2}=0$, in the Fisher approximation
the cut $\beta^{2}\ge0$ does not alter the confidence regions. Also,
we choose $\alpha=0.2$ as fiducial, but this should not have an important
impact since the tracking regime lasts very shortly; we tested also
the case $\alpha=0.1$. The other fiducial values are taken from WMAP7
\cite{Larson:2010gs} although since $\alpha\not=0$ our fiducial
model is not exactly $\Lambda$CDM.

All the derivatives of the Fisher matrix are performed numerically
by evaluating the spectra at two values $\theta_{i}(1\pm\varepsilon)$
where $\theta_{i}$ is a cosmological parameter (for parameters whose
fiducial is zero we use $\theta_{i}\pm\varepsilon$). Here we chose
$\varepsilon=0.03$ but we also tested that for $\varepsilon=0.06$
the final constraints change by at most 20\%.

\begin{center}
\begin{table}
\begin{centering}
\begin{tabular}{ll}
\hline 
\begin{minipage}[c]{100pt}%
\flushleft Parameter %
\end{minipage} & %
\begin{minipage}[c]{100pt}%
\flushleft Value %
\end{minipage}\tabularnewline
\hline 
$\Omega_{{\rm c}}h^{2}$  & 0.1116 \tabularnewline
$H_{0}$  & 70.3 km s$^{-1}$ Mpc$^{-1}$\tabularnewline
$\Omega_{{\rm b}}h^{2}$  & 0.02227 \tabularnewline
$n_{s}$  & 0.966\tabularnewline
$\beta^{2}$  & 0 \tabularnewline
$\alpha$  & 0.2 \tabularnewline
$A$  & $2.42\times10^{-9}$ \tabularnewline
\hline 
\end{tabular}
\par\end{centering}

\caption{Cosmological fiducial model: parameters are consistent with the WMAP
7 year results for a $\Lambda$CDM cosmology \citep{lambdawebsite, Larson:2010gs}
plus a coupling.}

\label{cosmological_parameters} 
\end{table}

\par\end{center}

\section{Results from CMB data}

\begin{figure}
\includegraphics[scale=0.7]{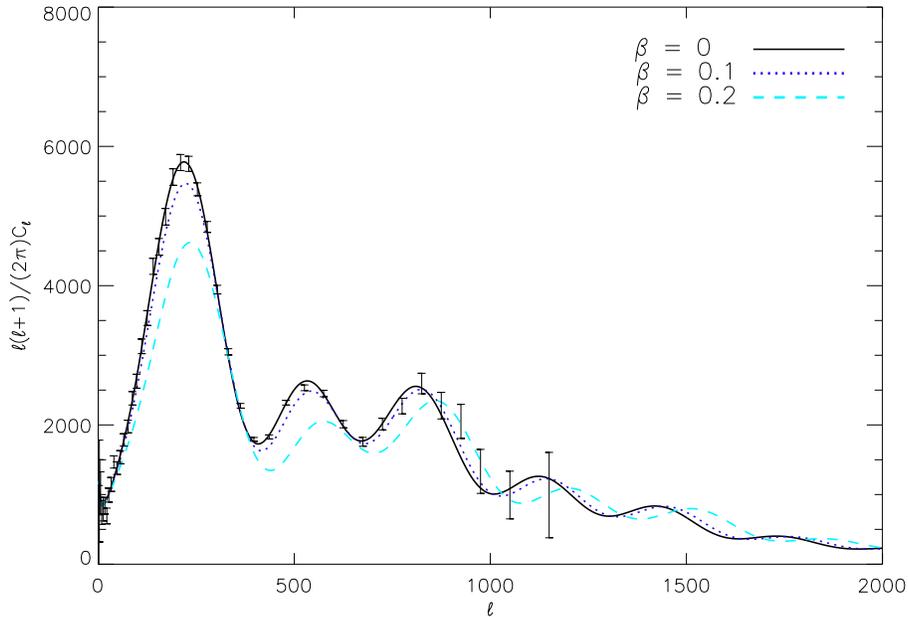}

\caption{\label{fig:CMB-for-three}CMB unlensed TT temperature spectra for
three values of $\beta$. Data are taken from WMAP7 \cite{Larson:2010gs}.}
\end{figure}

The coupling has two main effects on the CMB: 1) it moves the position
of the acoustic peaks to larger $\ell$'s due to the increase in the
last scattering surface distance (sometimes called projection effect,
\citep{Pettorino:2008ez} and references therein); 2) it reduces the
ratio of baryons to dark matter at decoupling with respect to its
present value, since coupled dark matter dilute faster than in an
uncoupled model. Both effects are clearly visible in Fig. (\ref{fig:CMB-for-three})
for some values of $\beta$.

We use the Fisher Matrix method described in section \ref{sec:methods}
to predict confidence contours for the cosmological parameter set
$\Theta\equiv\{\beta^{2},\alpha,\Omega_{c},h,\Omega_{b},n_{s},\log A\}$,
where $A$ is the spectrum normalization, around the chosen fiducial
model. The fiducial value of $A$ is chosen to be $A=2.42\times10^{-9}$,
as in WMAP7 best fit for $\Lambda$CDM \cite{lambdawebsite,Larson:2010gs}.
For our CMB analysis, we use Planck satellite \cite{planck} specifications
and include three frequency channels, each characterized by the experimental
specifications illustrated in Tab.\ref{tab:planck}. For each channel
we specify the FWHM (Full Width at Half Maximum) of the beam when
assuming a Gaussian profile and the temperature sensitivity $\sigma_{T}$.
The polarization sensitivity is given by $\sigma_{P}=\sqrt{2}\sigma_{T}$.
To each theoretical $C_{l}$ spectrum, we add a noise spectrum given
by: 
\begin{equation}
N_{l}=(\theta\sigma)^{2}\exp(l(l+1)/{l_{b}}^{2})
\end{equation}
 where $l_{b}$ is given by $l_{b}\equiv\sqrt{8\ln{2}}/\theta$ and
$\theta$ is the FWHM.

We obtained the CMB spectra by modifying the CAMB code \cite{Lewis:1999bs}
including the coupling; the output have been compared to an independent
code \cite{amendola_quercellini_2003} that is build on CMBFAST and
the agreement was better than 1$\%$. The initial conditions needed
to obtain the desired present values of the cosmological parameters
must be found by trial and error, through an iterative routine.

\label{results}

\begin{table}
\begin{tabular}{llll}
\hline 
\begin{minipage}[c]{80pt}%
\flushleft Planck %
\end{minipage} & %
\begin{minipage}[c]{80pt}%
\flushleft Channel {[}GHz{]} %
\end{minipage} & %
\begin{minipage}[c]{70pt}%
\flushleft FWHM %
\end{minipage} & %
\begin{minipage}[c]{70pt}%
\flushleft $\sigma_{T}[\mu K/K]$ %
\end{minipage}\tabularnewline
\hline 
$f_{\mathrm{sky}}$ = 0.85  & 70  & 14'  & 4.7 \tabularnewline
 & 100  & 10'  & 2.5 \tabularnewline
 & 143  & 7.1'  & 2.2 \tabularnewline
\hline 
\end{tabular}\caption{Planck experimental specifications \cite{planck}. FWHM is the Full
Width at Half Maximum of the beam, assuming a Gaussian profile and
expressed in arc-minutes. The polarization sensitivity is $\sigma_{P}=\sqrt{2}\sigma_{T}$.}

\label{tab:planck} 
\end{table}

\begin{figure*}
\centering \includegraphics[width=15cm]{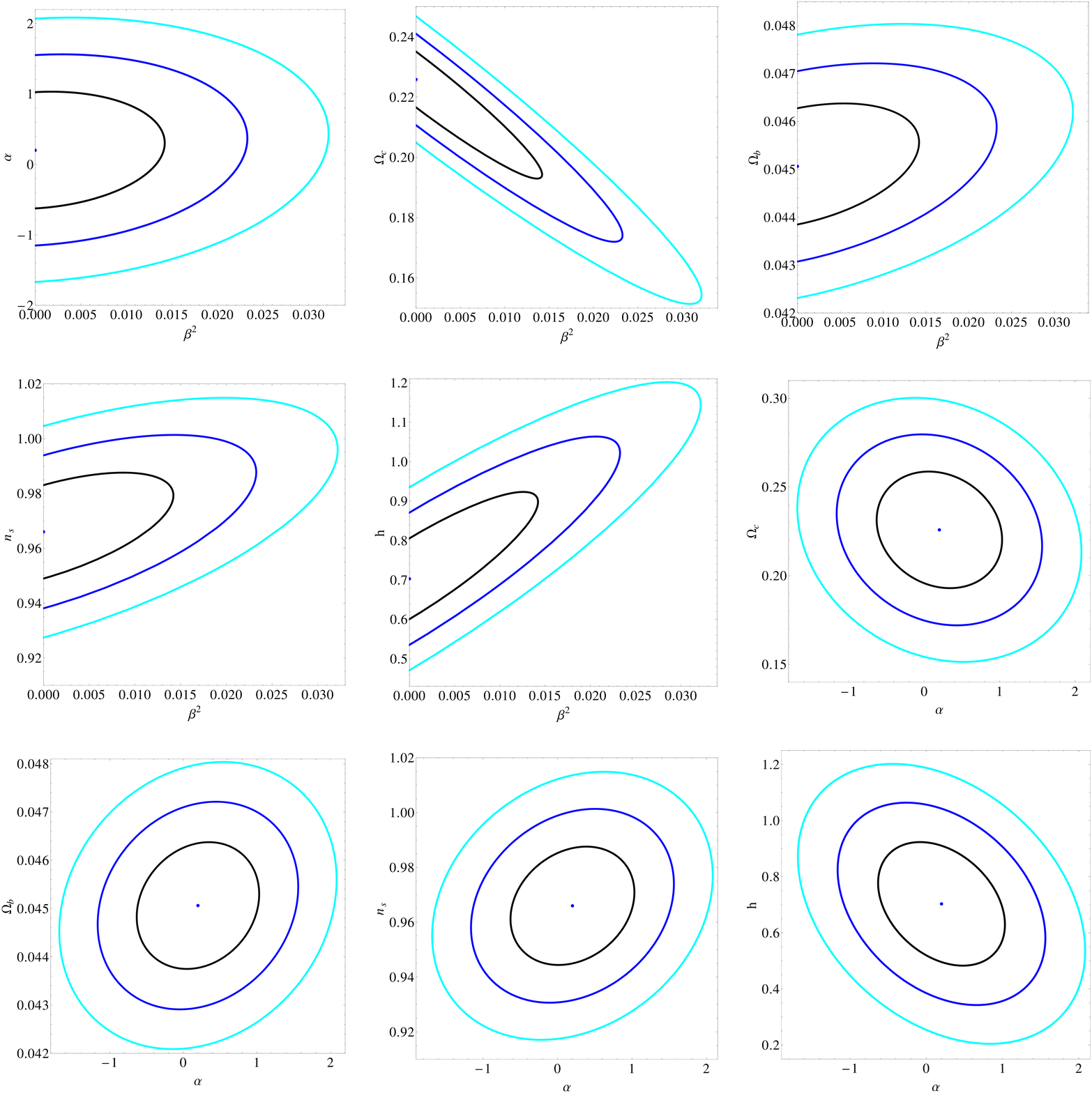}\caption{{\small Predicted confidence contours for the cosmological parameter
set $\Theta\equiv\{\beta^{2},\alpha,\Omega_{c},h,\Omega_{b},n_{s},\log A\}$
using CMB Planck specifications.}}

{\small \label{fig:cmb-cont} } 
\end{figure*}

The main results are in Fig.(\ref{fig:cmb-cont}), obtained marginalizing
over all parameters except those in the axes. Notice the sharp degeneracy
$\beta^{2}$ vs. $h$, in agreement with \cite{amendola_quercellini_2003},
and with $\Omega_{c}$ ans $n_{s}$. This can be understood in the
context of coupled dark energy. With respect to a cosmological constant
model, coupled dark matter dilutes more rapidly so that $\rho_{m}$
was higher in the past, leading to a faster expansion and a consequent
smaller size of the sound horizon at the last scattering surface,
not fully compensated by the faster expansion after decoupling; this
moves the acoustic peak towards larger $\ell$s, an effect which is
degenerated with an increase of $h$ (Fig.(\ref{fig:cmb-cont}), central
panel). On the other hand, a stronger coupling decreases the peak
amplitudes because $\rho_{b}/\rho_{m}$ at decoupling gets smaller:
this effect is compensated by lower values of $\Omega_{c}$ or by
higher values of the spectral index $n_{s}$. In Table (\ref{tab:cmb})
we report the fully marginalized 1$\sigma$ errors for all parameters;
for convenience of comparison we also report here the results of the
next sections.

\begin{figure*}
\centering \includegraphics[width=16cm]{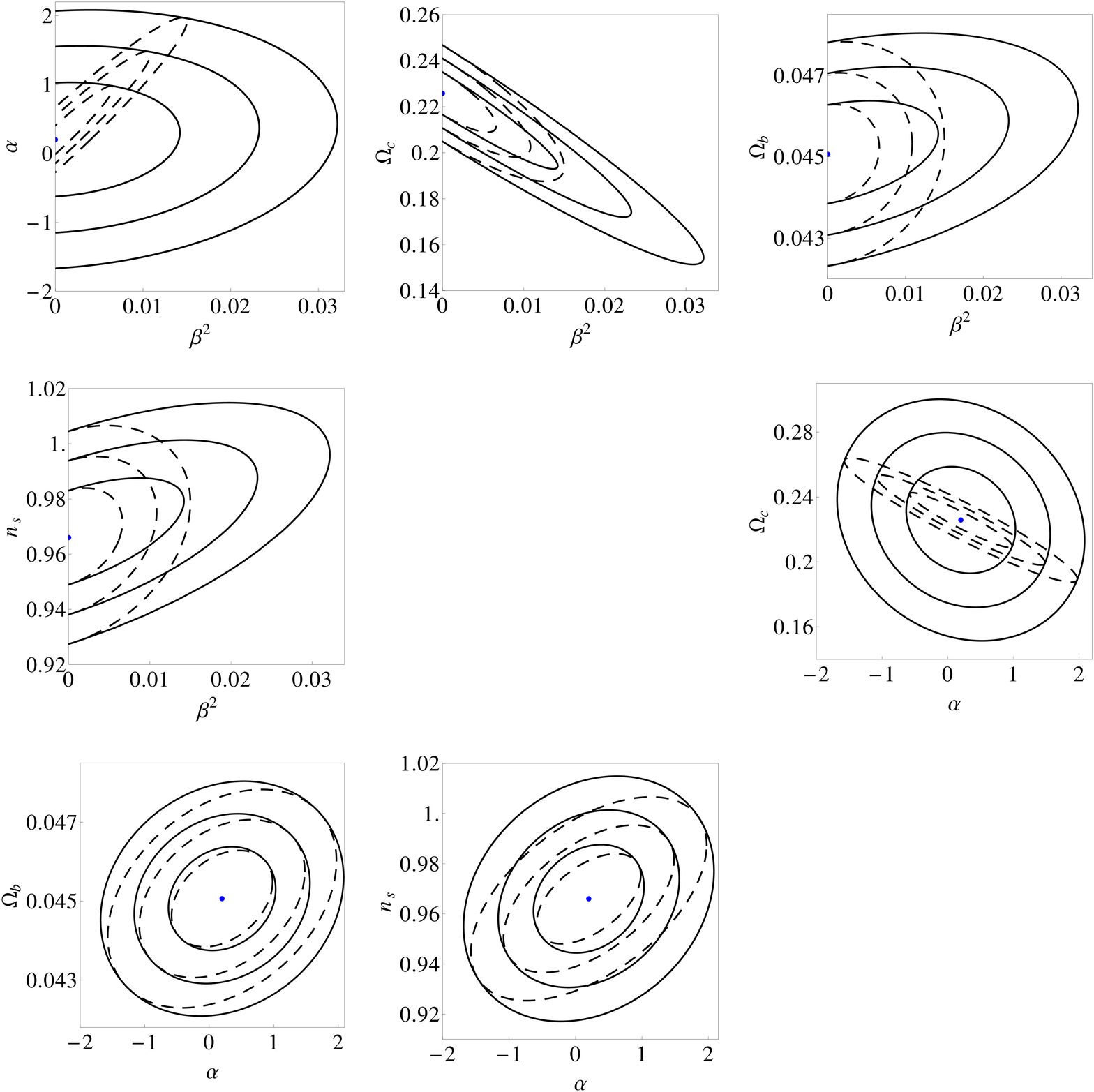}\caption{{\small Predicted confidence contours for the cosmological parameter
set $\Theta\equiv\{\beta^{2},\alpha,\Omega_{c},h,\Omega_{b},n_{s},\log A\}$
(in black, equal to Fig. (\ref{fig:cmb-cont})) using CMB Planck specifications.
Overplotted in dashed are the predicted confidence contours for the
cosmological parameter set
$\{\beta^{2},\alpha,\Omega_{c},\Omega_{b},n_{s},\log A\}$,
with $h$ fixed to the reference value.}}

{\small \label{fig:fish-cont_h} } 
\end{figure*}

\begin{figure*}
\centering \includegraphics[bb=0bp 350bp 1500bp 1150bp,clip,width=15cm]{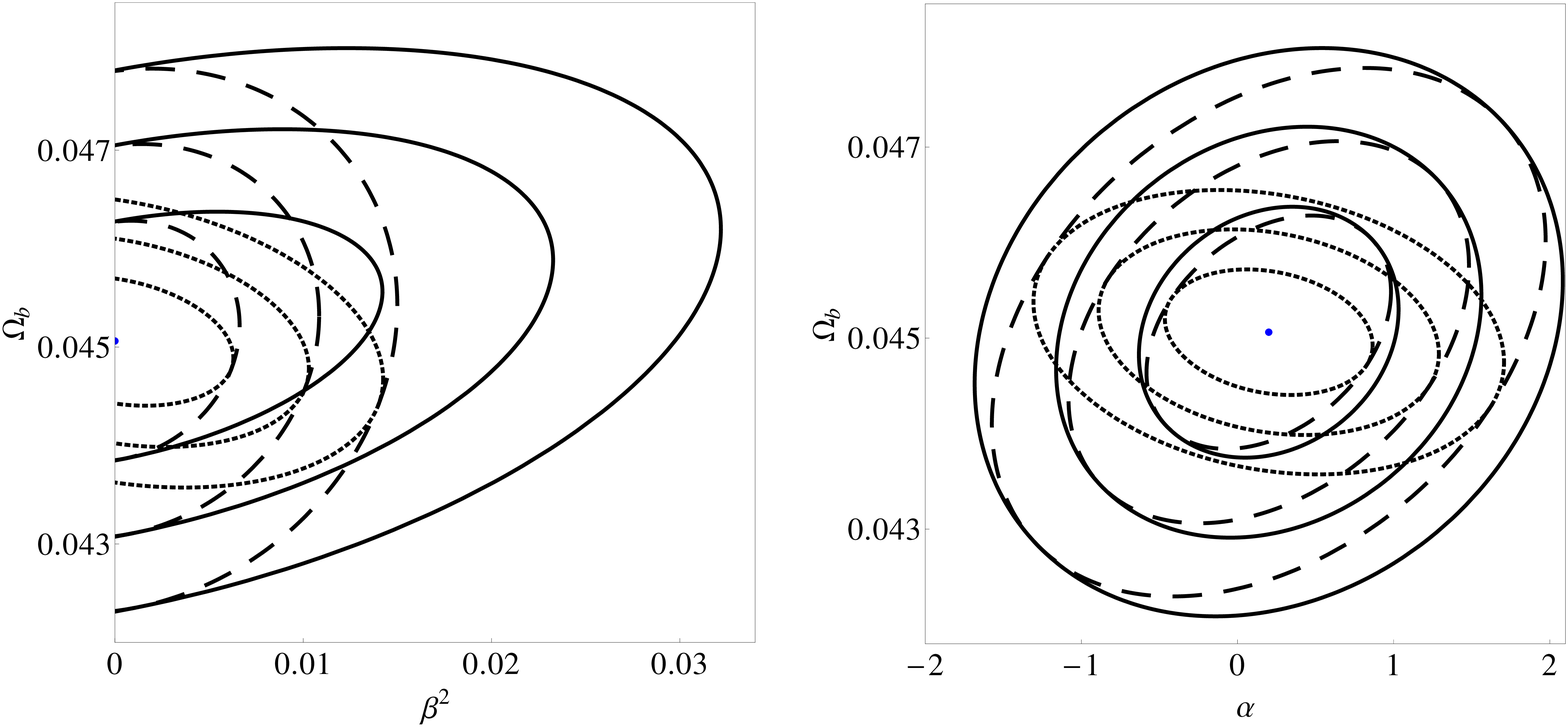}\caption{{\small Predicted confidence contours for the cosmological parameter
set $\Theta\equiv\{\beta^{2},\alpha,\Omega_{c},h,\Omega_{b},n_{s},\log A\}$
(in black, equal to fig.(\ref{fig:fish-cont_h})) using CMB Planck
specifications. Overplotted are the predicted confidence contours
for the cosmological parameter set $\{\beta^{2},\alpha,\Omega_{c},\Omega_{b},n_{s},\log A\}$
(dashed contours, with $h$ fixed to the reference values) and the cosmological
parameter set $\{\beta^{2},\alpha,\Omega_{c},\Omega_{b},\log A\}$
(dotted contours, with $h$ and $n_{s}$ fixed to the reference values).}}

{\small \label{fig:fish-cont_h_ns} } 
\end{figure*}

In Fig. (\ref{fig:fish-cont_h}) instead we fix $h$, i.e. we assume
that its fiducial value has a negligible error. If we fix $h$ and
$n_{s}$ we get Fig. (\ref{fig:fish-cont_h_ns}); the other contours
do not improve with respect to fixing $h$ only (see Table \ref{tab:cmb-fixed-cases}).
The significant improvement on the errors shows that there is much
to gain by reducing the errors on $h,n_{s}$. The error on $\beta^{2}$
goes from 0.009 (fully marginalized) to 0.004 (fixing both $h$ and
$n_{s}$), with a consequent better estimation of $\Omega_{c}$.

\begin{table*}
\begin{tabular}{lllll}
\begin{minipage}[c]{70pt}%
\flushleft Parameter %
\end{minipage} & %
\begin{minipage}[c]{70pt}%
\flushleft $\sigma_{i}$ CMB %
\end{minipage} & %
\begin{minipage}[c]{70pt}%
\flushleft $\sigma_{i}$ $P(k)$ %
\end{minipage} & %
\begin{minipage}[c]{70pt}%
\flushleft $\sigma_{i}$ WL %
\end{minipage} & \tabularnewline
\hline 
$\beta^{2}$  & 0.0094  & 0.0015  & 0.012  & \tabularnewline
$\alpha$  & 0.55  & 0.12  & 0.083  & \tabularnewline
$\Omega_{c}$  & 0.022  & 0.010  & 0.012  & \tabularnewline
$h$  & 0.15  & 0.036  & 0.039  & \tabularnewline
$\Omega_{b}$  & 0.00087  & 0.0022  & 0.010  & \tabularnewline
$n_{s}$  & 0.014  & 0.034  & 0.026  & \tabularnewline
$\sigma_{8}$  & -  & 0.0084  & 0.024  & \tabularnewline
$\log A$  & 0.0077  & -  & -  & \tabularnewline
\hline 
\label{tab:cmb}  &  &  &  & \tabularnewline
\end{tabular}

\caption{1-$\sigma$ errors for the set $\Theta\equiv\{\beta^{2},\alpha,\Omega_{c},h,\Omega_{b},n_{s},\sigma_{8},\log A\}$
of cosmological parameters, using CMB data, $P(k)$ and Weak Lensing
(WL). Here and in all the following tables the errors are fully marginalized. }
\end{table*}

\begin{table*}
\begin{tabular}{lllll}
\begin{minipage}[c]{100pt}%
\flushleft Parameter %
\end{minipage} & %
\begin{minipage}[c]{100pt}%
\flushleft $\sigma_{i}$ CMB %
\end{minipage} & %
\begin{minipage}[c]{100pt}%
\flushleft $\sigma_{i}$ CMB($\bar{h}$) %
\end{minipage} & %
\begin{minipage}[c]{100pt}%
\flushleft $\sigma_{i}$ CMB($\bar{h},\bar{n}_{s}$) %
\end{minipage} & \tabularnewline
\hline 
$\beta^{2}$  & 0.0094  & 0.0044  & 0.0041  & \tabularnewline
$\alpha$  & 0.55  & 0.52  & 0.44  & \tabularnewline
$\Omega_{c}$  & 0.022  & 0.011  & 0.0079  & \tabularnewline
$h$  & 0.15  & -  & -  & \tabularnewline
$\Omega_{b}$  & 0.00087  & 0.00081  & 0.00043  & \tabularnewline
$n_{s}$  & 0.014  & 0.012  & -  & \tabularnewline
$\sigma_{8}$  & 0  & 0  & 0  & \tabularnewline
$log(A)$  & 0.0077  & 0.0074  & 0.0057  & \tabularnewline
\hline 
\end{tabular}\label{tab:cmb-fixed-cases} \caption{1-$\sigma$ errors using CMB for the set $\Theta\equiv\{\beta^{2},\alpha,\Omega_{c},h,\Omega_{b},n_{s},\sigma_{8},\log(A)\}$
(left column), $\Theta\equiv\{\beta^{2},\alpha,\Omega_{c},\Omega_{b},n_{s},\sigma_{8},\log(A)\}$
(middle column, $h$ fixed to the reference value), $\Theta\equiv\{\beta^{2},\alpha,\Omega_{c},h,\Omega_{b},\sigma_{8},\log(A)\}$
(right column, $h$ and $n_{s}$ both fixed to their reference values).}
\end{table*}

\section{Results for the galaxy power spectrum}

\label{sec:p(k)}

\begin{figure*}
\centering \includegraphics[width=10cm]{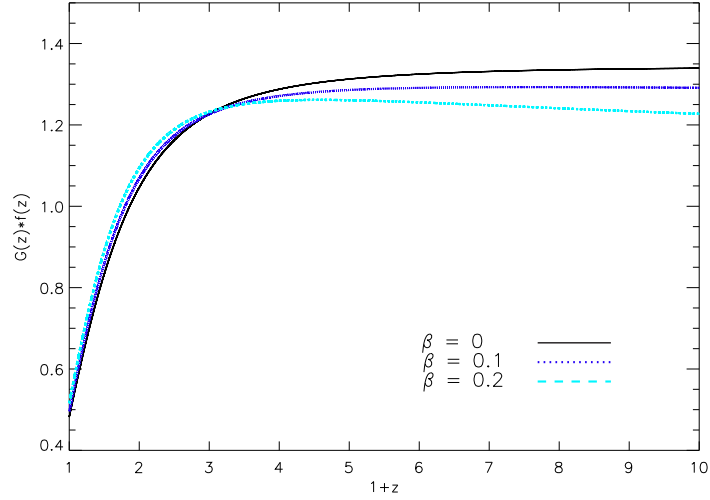}\caption{{\small The combination $G(z)f(z)$ that appears as the amplitude
of the power spectrum, for various values of $\beta$. $G(z)$ is
normalized to unity today.}}

{\small \label{fig:-g(z)}} 
\end{figure*}

\begin{figure*}
\centering \includegraphics[width=10cm]{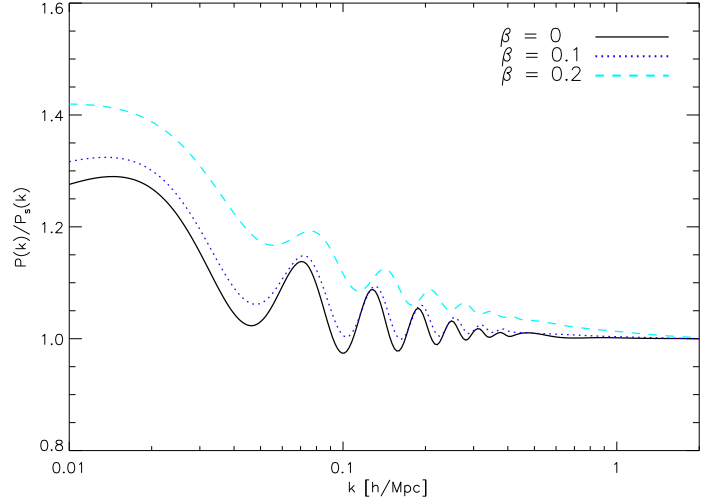}\caption{{\small BAO oscillations in $P(k)$ for three values of $\beta$.
The curves show the ratio $P(k)/P_{s}(k)$, where $P_{s}(k)$ is the
corresponding smooth power spectrum.}}

{\small \label{fig:bao} } 
\end{figure*}

The main effects of the coupling on the matter power spectrum are
the shift in the matter-radiation equality, the change in location
and amplitude of the BAO and the speed up in the perturbation growth.
As already noted, since dark matter dilutes faster for larger $|\beta|$,
there is more dark matter in the past than in uncoupled cases and
therefore the equality moves to higher redshifts. This implies that
the wavelenghts at which perturbations reenter during radiation domination
are smaller and thus the power spectrum turnaround moves to smaller
scale. Just as for the CMB, the acoustic peaks also move to smaller
scales and their amplitude is reduced. These effects are clearly visible
in Figs. (\ref{fig:-g(z)},\ref{fig:bao}). The difference in the
parameter growth function is plotted in Fig.(\ref{fig:-g(z)}). In
Fig.(\ref{fig:bao}) we make manifest the effect of the coupling on
baryonic acoustic oscillations. For each $\beta$ we plot the ratio
between the power spectrum and its smooth spectrum (obtained reducing
the amount of baryons in the CAMB code). The ratios are then normalized
to high momenta $k$. As we can see the coupling affects both the
amplitude and position of the BAO peaks, shifting them to higher momenta.
The effect of $\beta$ on $P(k,z)$ is therefore quite strong since
it occurs simultaneously on the wiggle position and amplitude, on
the broad spectrum shape, and on the growth. This results in a tight
constraint on the coupling parameter.

Following~\cite{2003ApJ...598..720S}, we build a function that provides
the observed linear spectrum at all redshifts and for all cosmological
parameters: 
\begin{equation}
P_{r,obs}(k_{r},\mu_{r};z)=P_{s}(z)+\frac{D_{r}^{2}(z)H(z)}{D^{2}(z)H_{r}(z)}G^{2}(z)b^{2}(z)\sigma_{8}^{2}(1+\beta_{d}(z)\mu^{2})^{2}P(k,z=0)\,,
\end{equation}
 where $G(z)$ is the growth factor, $b(z)$ is the bias, $\beta_{d}(z)$
is the redshift-distortion factor, $P(k,z=0)$ is the undistorted
linear matter spectrum at $z=0$ normalized to unity, $\mu$ is the
direction cosine, $D$ and $H$ are the angular diameter distance
and the Hubble rate at the shell redshift $z$, respectively, and
$P_{s}(z)$ is the $z$- dependent shot-noise correction, to be marginalized
over in every redshift bin. The subscript $r$ (for `reference') indicates
quantities calculated in the fiducial model. In linear theory we have
$\beta_{d}(z)=f(z)/b(z)$ where $f(z)=d\log G/d\log a$ is the growth
rate.

In this case, the Fisher matrix for every redshift bin shell is \cite{2003ApJ...598..720S},
\begin{equation}
F_{ij}=\frac{1}{8\pi^{2}}\int_{-1}^{+1}\dd\mu\int_{k_{min}}^{k_{max}}k^{2}\dd k\,\frac{\partial\ln P_{obs}(k,\mu)}{\partial\theta_{i}}\frac{\partial\ln P_{obs}(k,\mu)}{\partial\theta_{j}}\left[\frac{n(z)P_{obs}(k,\mu)}{n(z)P_{obs}(k,\mu)+1}\right]^{2}V_{s}\,.\label{eq:pkfm}
\end{equation}
 where $n(z)$ is the galaxy number density at redshift $z$ and $V_{s}$
is the survey volume of the redshift shell. The power spectrum at
$z=0$ and the functions $G(z),H(z),D(z)$ are all obtained numerically
by solving the background and perturbation equations of the system
for each value of the parameters and the derivatives in $F_{ij}$
are evaluated numerically. We follow \cite{Amendola2005,diporto10,Wang2010}
and instead of marginalizing over $G$ and $f$ we express them in
function of the parameters; in particular we form the combination
$g(z)\equiv Gf\sigma_{8}$ for each redshift bin and write 
\[
G^{2}(z)b^{2}(z)\sigma_{8}^{2}(1+\beta_{d}(z)\mu^{2})^{2}P(k,z=0)=\frac{g(z)^{2}}{\beta_{d}(z)^{2}}(1+\beta_{d}(z)\mu^{2})^{2}P(k,z=0)
\]
 and we marginalize over $\beta_{d}(z)$ assuming an independent parameter
$\beta_{d}$ for every redshift bin. The prescription for $k_{max}$
is that the variance in cells $\sigma^{2}(k_{max},z)=0.25$, resulting
in a $k_{max}=0.16h/$Mpc for the first shell at $z=0.6$. This conservative
cut of the higher momenta allows us to discard the problem of the
non-linear correction.

Here and in the next section we have a new parameter, namely $\sigma_{8}$,
whose fiducial is fixed to 0.8. In principle, the parameter $\log A$
employed in the CMB section is related to $\sigma_{8}$ and therefore
we should not count them separately. However, the two normalizations
are taken at very different epochs and we are being conservative by
assuming them to be independent.

\begin{figure*}
\centering \includegraphics[width=15cm]{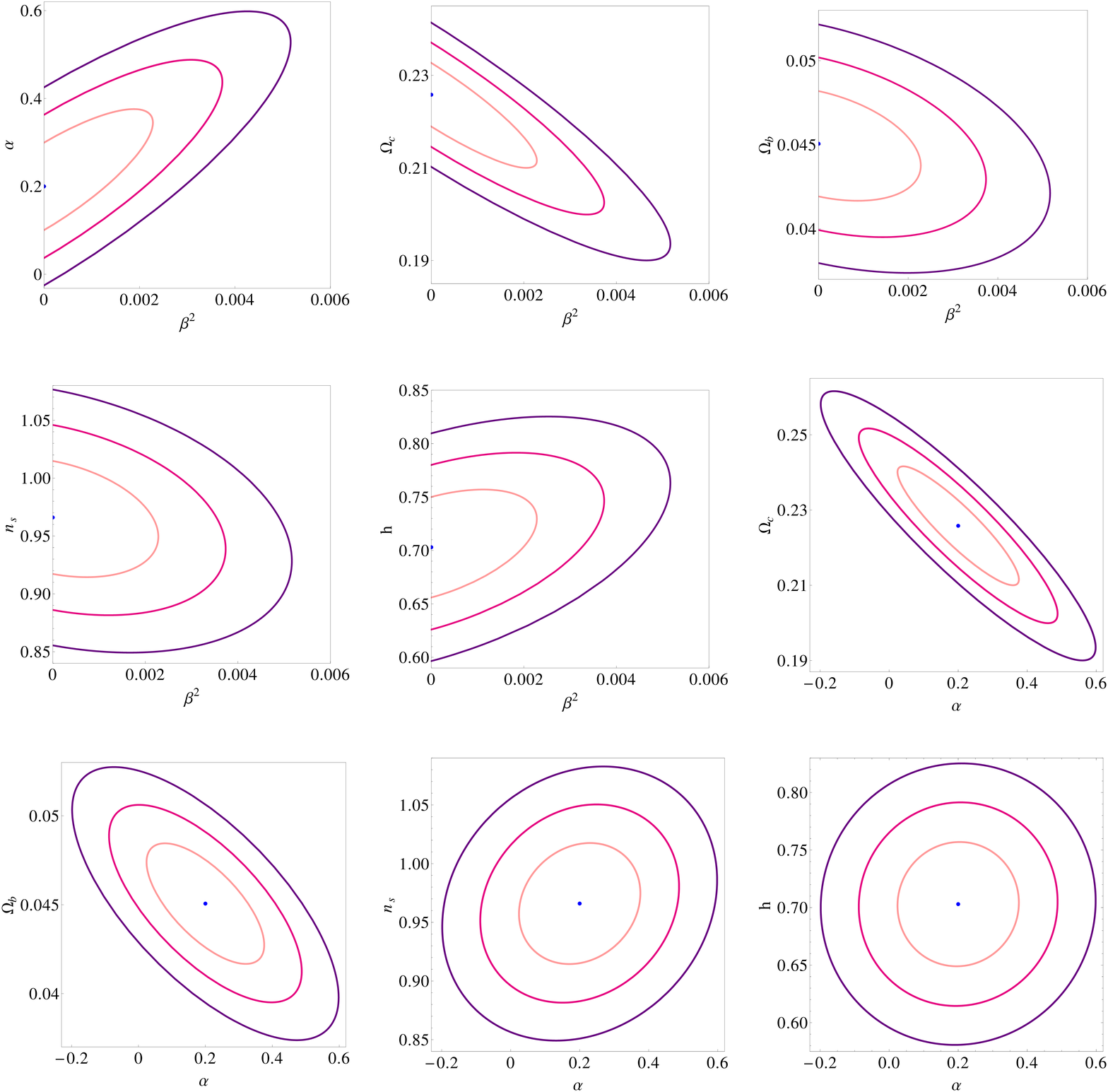}\caption{{\small Predicted confidence contours for the cosmological parameter
set $\Theta\equiv\{\beta^{2},\alpha,\Omega_{c},h,\Omega_{b},n_{s},\sigma_{8}\}$
using the galaxy power spectrum.}}

{\small \label{fig:bao-cont} } 
\end{figure*}

We use specifications for a Euclid-like mission based on the Euclid
Red Book \cite{euclidredbook}. The fiducial bias $b(z)$ follows
\cite{Orsi:2009mj}; we adopt redshift bins of $\Delta z=0.1$ between
$0.5<z<2$ and a sky coverage of 15,000 sq. deg. \cite{euclidredbook}.
The expected number of $H\alpha$ emitters per $\deg^{2}$ between
$z-dz/2$ and $z+dz/2$ is computed for $H\alpha$ flux $F_{H\alpha}=4.00\times10^{-16}erg/s/cm^{2}$
according to \cite{geach_etal_2009}. 
 In Fig. (\ref{fig:bao-cont}) we show the expected confidence regions
from galaxy power spectra.

The Euclid-like $P(k)$ probe gives in general better constraints
than Planck (and also WL, see below), as pointed out in \cite{euclidredbook}.
This applies also to the coupling parameter $\beta$: we find indeed
$\beta^{2}<0.0015$, a constraint six times stronger than for Planck
CMB.

\section{Results for Weak lensing}

\label{sec:wl} An excellent probe for dark energy parameters that
complements the aforementioned techniques is weak lensing (see \cite{Bartelmann:1999yn}
for an extensive treatment), due to the sensitivity with respect to
the growth of structure and equal treatment of dark and baryonic matter.
The weak lensing power spectrum depending on the multipole $\ell$
can be written in terms of the matter power spectrum as \cite{Kaiser1992}
\begin{equation}
P(\ell)=\frac{9}{4}\int_{0}^{\infty}\mathrm{d}z\frac{W^{2}(z)H^{3}(z)\Omega_{m}^{2}(z)}{(1+z)^{4}}P_{m}\left(\frac{\ell}{\pi r(z)}\right),\label{eq:convergenceps}
\end{equation}
 where 
\begin{equation}
W(z)=\int_{z}^{\infty}\mathrm{d}\tilde{z}\left(1-\frac{r(z)}{r(\tilde{z})}\right)n(\tilde{z})\label{eq:window}
\end{equation}
 is the window function and 
\begin{equation}
n(z)=z^{2}\exp\left(-(z/z_{0})^{3/2}\right)
\end{equation}
 the normalized galaxy distribution function \cite{Amara2007}. Note
that $z_{0}$ is related to the median redshift via $z_{\mathrm{med}}\approx1.412z_{0}$.

\begin{figure}
\includegraphics{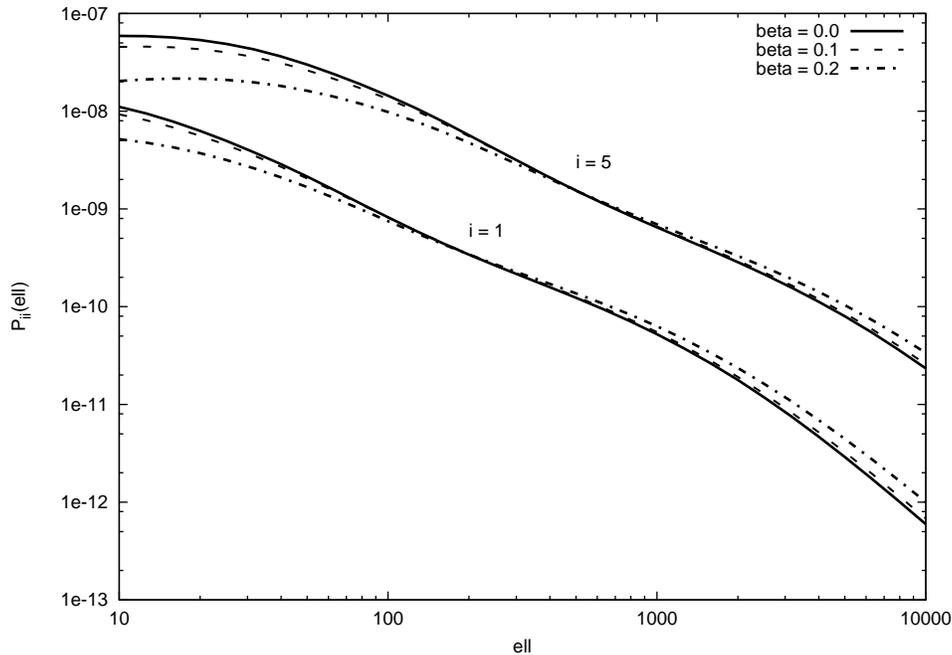} \caption{\label{fig:The-convergence-power}The convergence power spectra $P_{11}$
and $P_{55}$ for three values of $\beta$.}
\end{figure}

In order to extract as much information from the cosmic shear as possible,
we will also employ the so-called weak lensing tomography \cite{Hu1999},
where we form $\mathcal{N}=5$ redshift bins and compute the cross-correlation
and power spectrum of the shear field. With this, eq.~(\ref{eq:convergenceps})
becomes~\cite{Hu1999} 
\begin{equation}
P_{ij}(\ell)=\frac{9}{4}\int_{0}^{\infty}\mathrm{d}z\frac{W_{i}(z)W_{j}(z)H^{3}(z)\Omega_{m}^{2}(z)}{(1+z)^{4}}P_{m}\left(\frac{\ell}{\pi r(z)}\right),\label{eq:pstomog}
\end{equation}
 where $W_{i}(z)$ now depends on $n_{i}(z)$, the galaxy distribution
in the $i$-th redshift bin. The binned galaxy distribution is normalized
to unity and has been convolved with a Gaussian to account for photometric
redshift errors $\Delta_{z}(1+z)$~, i.e. 
\begin{equation}
n_{i}(z)=A_{i}\int\limits _{i\text{-th bin}}n(\tilde{z})\exp\left(\frac{-(\tilde{z}-z)^{2}}{2(\Delta_{z}(1+\hat{z}_{i}))^{2}}\right)\mathrm{d}\tilde{z}
\end{equation}
 where $\hat{z}_{i}$ is the center of the $i$-th redshift bin and
$A_{i}$ a normalization factor. In Fig. (\ref{fig:The-convergence-power})
we show the convergence power spectrum for three values of $\beta$.
The behavior is not trivial since the spectra are a convolution of
background and perturbation evolution.

The Fisher matrix for the weak lensing power spectrum is then given
by \cite{Hu1999a} 
\begin{equation}
F_{\alpha\beta}=f_{\mathrm{sky}}\sum\limits _{\ell,i,j,k,m}\frac{(2\ell+1)\Delta\ell}{2}\frac{\partial P_{ij}(\ell)}{\partial\theta_{\alpha}}C_{jk}^{-1}\frac{\partial P_{km}(\ell)}{\partial\theta_{\beta}}C_{mi}^{-1}\label{eq:fm-wl}
\end{equation}
 with the covariance matrix 
\begin{equation}
C_{ij}=P_{ij}+\delta_{ij}\gamma_{\mathrm{int}}^{2}n_{i}^{-1},
\end{equation}
 where $\gamma_{\mathrm{int}}=0.22$ in the shot noise term is the
intrinsic galaxy ellipticity \cite{Amara2007} and 
\begin{equation}
n_{i}=3600\left(\frac{180}{\pi}\right)^{2}n_{\theta}/\mathcal{N}
\end{equation}
 with $n_{\theta}$ being the total number of galaxies per $\mathrm{arcmin}^{2}$,
assuming that the redshift bins have been chosen such that each contain
the same amount of galaxies. Since we consider multipoles up to $\ell_{\mathrm{max}}=5000$,
we need to apply non-linear corrections to the matter power spectrum,
for which we use the fitting formulae from Ref. \cite{Smith2003}.
It is clear that this is far from satisfactory since these non-linear
corrections are calibrated through $\Lambda$CDM $N$-body simulations
and should not be used outside these cases. However at the moment
there are no suitable analytical extensions to coupled models (for
a recent effort in this direction see e.g. \cite{Baldi_etal_2010})
so we cannot improve on this. Moreover, since our fiducial is indeed
almost a $\Lambda$CDM cosmology we can assume that the non-linear
correction does not introduce a large bias in our results.

\begin{figure*}
\centering \includegraphics[width=15cm]{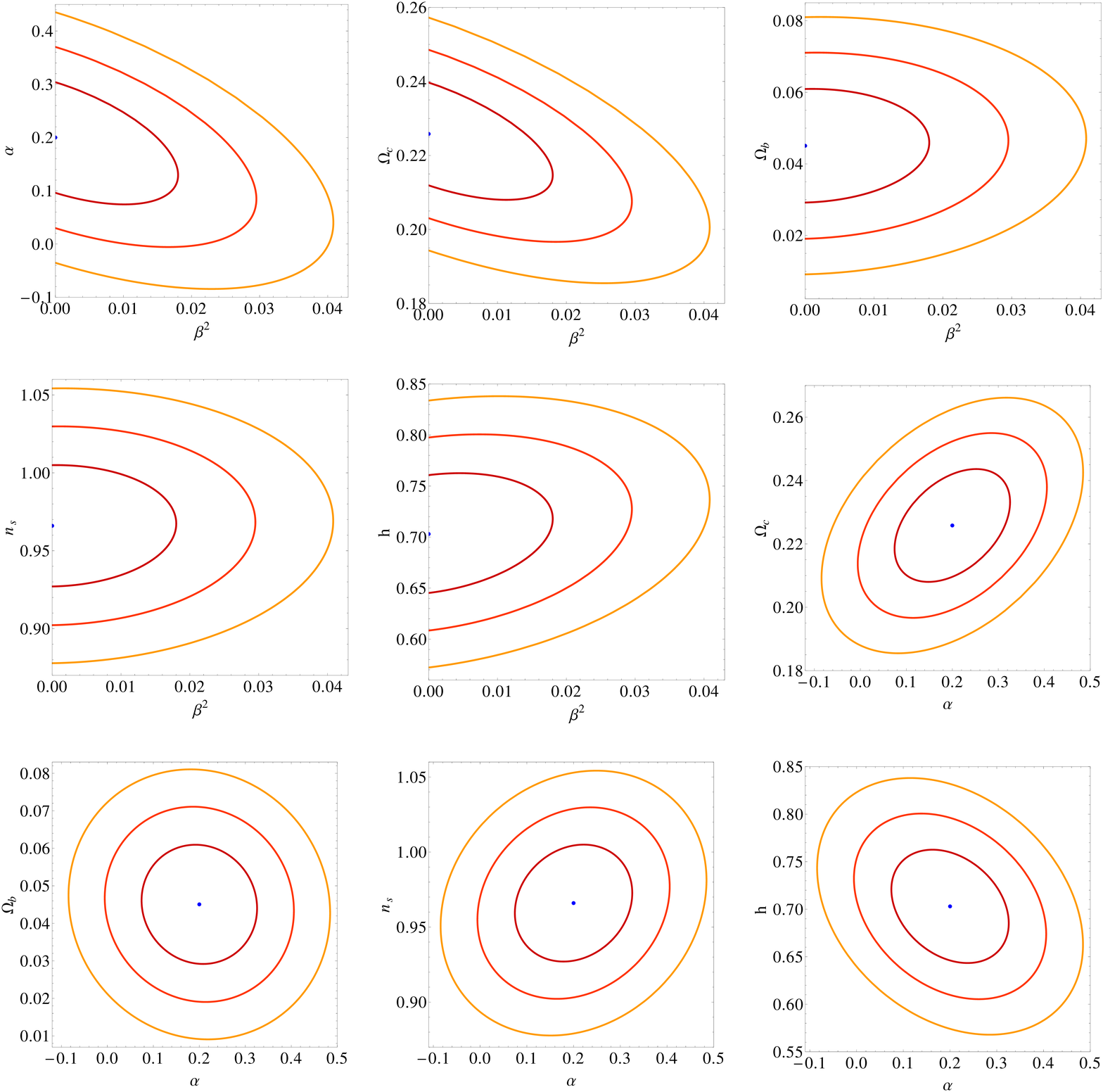}\caption{{\small Predicted confidence contours for the cosmological parameter
set $\Theta\equiv\{\beta^{2},\alpha,\Omega_{c},h,\Omega_{b},n_{s},\sigma_{8}\}$
using weak lensing Euclid-like specifications.}}

{\small \label{fig:wl-cont} } 
\end{figure*}

As before, we select a Euclid-like mission as our probe. We use the
specifications listed in Table~\ref{tab:euclid}, as taken from the
Euclid Red Book \cite{euclidredbook}. In Fig. (\ref{fig:wl-cont})
we show the marginalized confidence contours. The WL constraint on
$\beta^{2}$ is $0.012$, weaker than either CMB or $P(k)$. Interestingly,
however, the constraint on the potential slope $\alpha$ is the strongest
among the various probes.

\begin{table}
\begin{tabular}{lll}
\hline 
\begin{minipage}[c]{100pt}%
\flushleft Parameter %
\end{minipage} & %
\begin{minipage}[c]{100pt}%
\flushleft Value %
\end{minipage} & %
\begin{minipage}[c]{100pt}%
\flushleft Description %
\end{minipage}\tabularnewline
\hline 
$A$  & 15 000 deg${^{2}}$  & Survey area\tabularnewline
$n_{\theta}$  & 30  & Galaxies per arcmin$^{2}$\tabularnewline
$\ell_{\mathrm{max}}$  & $5000$  & Maximum multipole\tabularnewline
$z_{\mathrm{med}}$  & 0.9  & Median redshift\tabularnewline
$\Delta_{z}$  & 0.05  & Photometric redshift error\tabularnewline
\hline 
\end{tabular}\caption{Mission goals for a Euclid-like project. \cite{euclidredbook}}

\label{tab:euclid} 
\end{table}

\section{Combining CMB, P(k) and WL}

In this Section we finally combine all the results of the various
probes. The combination of $P(k)$ and WL is not totally accurate
since the two probes are assumed to be independent while in fact they
observe the same field of galaxies and the lensing signal is correlated
with the density distribution. However, since as we have seen the
constraints from $P(k)$ are in general quite stronger, we expect
that an improved treatment will not change drastically our results.

\begin{figure*}
\centering \includegraphics[width=15cm]{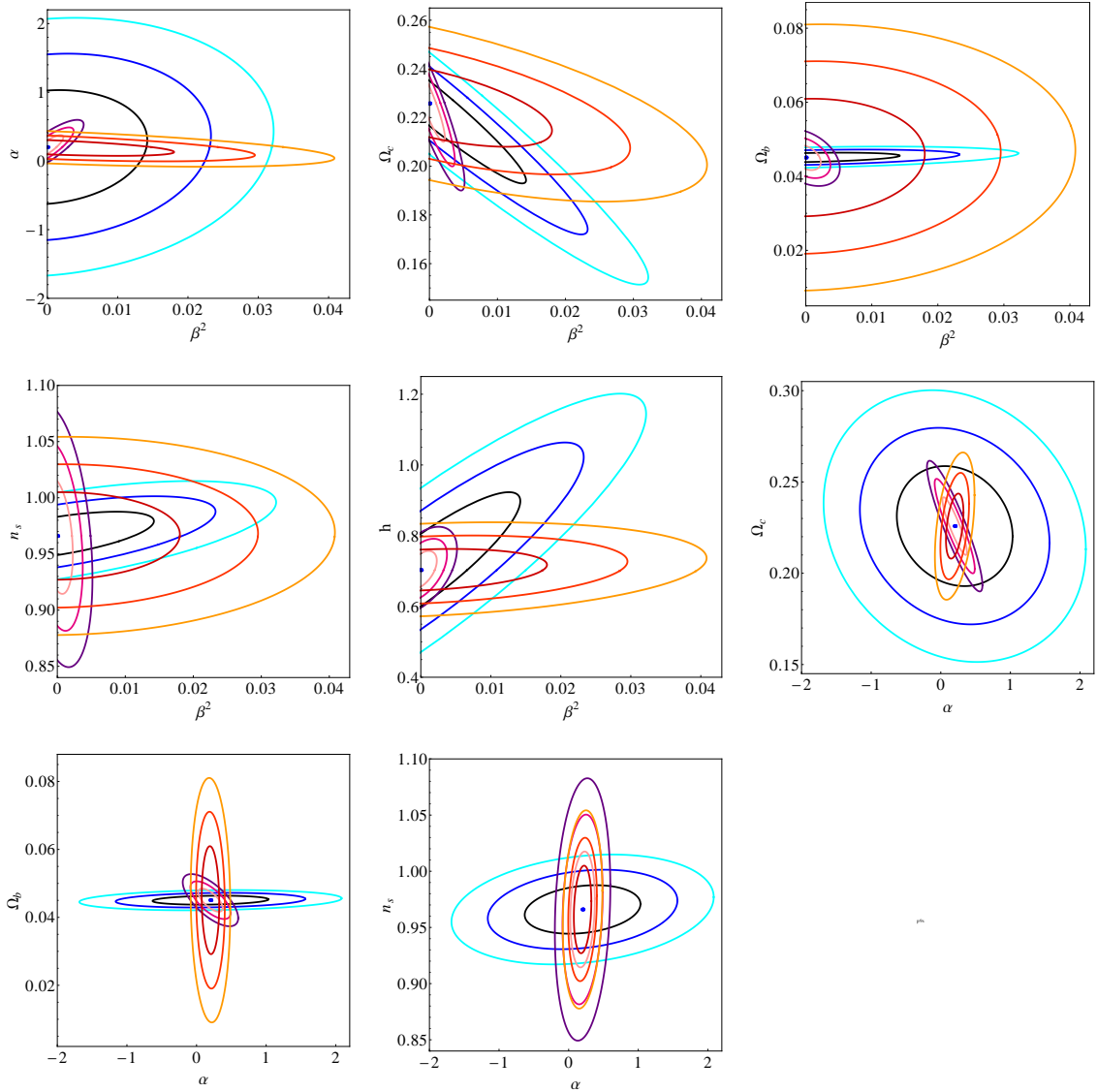}\caption{{\small Comparison among predicted confidence contours for the cosmological
parameter set $\Theta\equiv\{\beta^{2},\alpha,\Omega_{c},h,\Omega_{b},n_{s},\sigma_{8},log(A)\}$
using CMB (Planck, blue contours), $P(k)$ (pink-violet contours) and weak
lensing (orange-red contours) with Euclid-like specifications.}}

{\small \label{fig:cmbbaowl-cont} } 
\end{figure*}

\begin{figure*}
\centering \includegraphics[width=15cm]{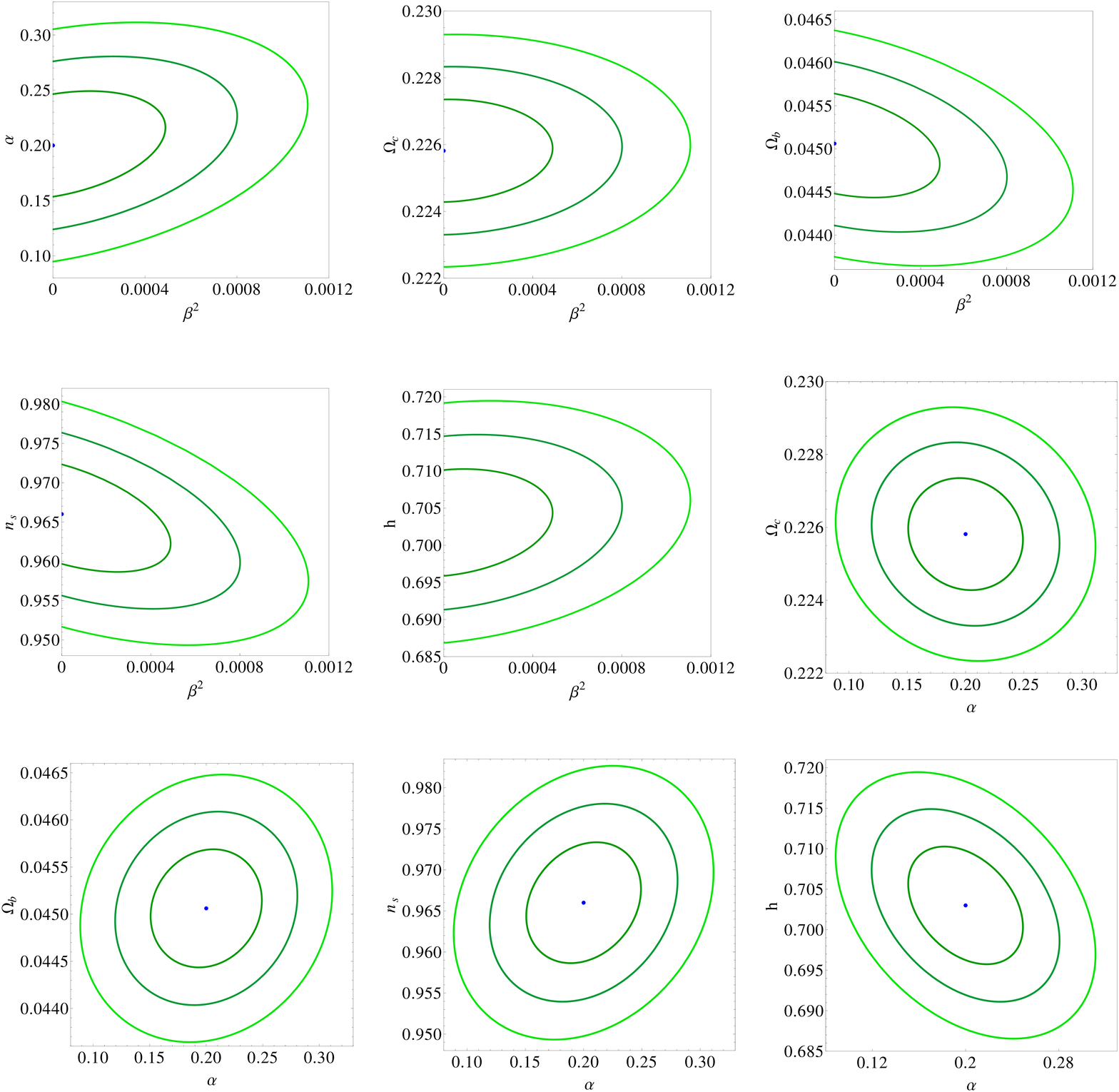}\caption{{\small Combined predicted confidence contours for the cosmological
parameter set $\Theta\equiv\{\beta^{2},\alpha,\Omega_{c},h,\Omega_{b},n_{s},\sigma_{8},\log(A)\}$
from CMB (Planck), $P(k)$ and weak lensing (Euclid-like) specifications.}}

{\small \label{fig:cmbbaowl-conb} } 
\end{figure*}

The combined Fisher confidence regions are plotted in Fig.(\ref{fig:cmbbaowl-cont})
and (\ref{fig:cmbbaowl-conb}) and the results are in Table (\ref{tab:combined}).
The main result is that future surveys can constrain the coupling
of dark energy to dark matter $\beta^{2}$ to less than $3\cdot10^{-4}$. 

\begin{table*}
\begin{tabular}{lllll}
\begin{minipage}[c]{100pt}%
\flushleft Parameter %
\end{minipage} & %
\begin{minipage}[c]{100pt}%
\flushleft $\sigma_{i}$ CMB+$P(k)$ %
\end{minipage} & %
\begin{minipage}[c]{100pt}%
\flushleft $\sigma_{i}$ CMB+$P(k)$+WL %
\end{minipage} &  & \tabularnewline
\hline 
$\beta^{2}$  & 0.00051  & 0.00032  &  & \tabularnewline
$\alpha$  & 0.055  & 0.032  &  & \tabularnewline
$\Omega_{c}$  & 0.0037  & 0.0010  &  & \tabularnewline
$h$  & 0.0080  & 0.0048  &  & \tabularnewline
$\Omega_{b}$  & 0.00047  & 0.00041  &  & \tabularnewline
$n_{s}$  & 0.0057  & 0.0049  &  & \tabularnewline
$\sigma_{8}$  & 0.0049  & 0.0036  &  & \tabularnewline
$\log(A)$  & 0.0051  & 0.0027  &  & \tabularnewline
\hline 
\end{tabular}\label{tab:combined} \caption{1-$\sigma$ errors for the set $\Theta\equiv\{\beta^{2},\alpha,\Omega_{c},h,\Omega_{b},n_{s}\,\sigma_{8},\log(A)\}$
of cosmological parameters, combining CMB+$P(k)$ (left column) and
CMB+$P(k)$+WL (right column).}
\end{table*}

We can ask whether a better knowledge of the parameters $\{\alpha,\Omega_{c},h,\Omega_{b},n_{s},\sigma_{8},\log(A)\}$,
obtained by independent future observations, can give us better constraints
on the coupling $\beta^{2}$. In Table \ref{tab:fixedparlist} we
list the 1-$\sigma$ error on $\beta^{2}$ obtained when we fix, one
after the other, each of the remaining parameters to their reference
value instead of marginalizing. $P_{1}$ corresponds to fixing one
parameter ($\alpha$) and marginalize over all the others, $P_{2}$
corresponds to fixing two parameters ($\alpha$ and $\Omega_{c}$)
and marginalize over all the others and so on until $P_{7}$ where
we estimate the error on $\beta^{2}$ when we fix all parameters.

The first column corresponds to the effect on CMB only. CMB alone
gains the most from a better knowledge of the parameters. In this
case the error on $\beta^{2}$ improves by almost one order of magnitude
when $\Omega_{c}$ is fixed (thanks to the degeneracy with this parameter)
and of almost two orders of magnitude when also $h$ and $\Omega_{b}$
are known. A better knowledge of $\alpha,\Omega_{c},h,\Omega_{b}$
makes CMB constraints alone comparable to the combination of all probes
(when all parameters are marginalized over).

$P(k)$ observations (second column) improve bounds by almost an order
of magnitude when $\alpha,\Omega_{c},h,\Omega_{b}$ have been fixed,
becoming even better than the ones obtained in the marginalized combination
of all probes. The third column shows weak lensing constraints, which
gain one order of magnitude when all parameters are known: note that
in this case the amplitude of perturbations $\sigma_{8}$ is still
relevant even when all previous parameters have already been fixed.

The last column combines all three probes. Interestingly, the combination
of CMB, power spectrum and weak lensing is already very good when
all parameters are marginalized over; no much gain is obtained, overall,
by a better knowledge of the other parameters. This is confirmed also
when looking at Table \ref{tab:fixedpar}. Here we show the errors
on $\beta^{2}$ when we have a better knowledge of only one other
parameter, which is here fixed to the reference value. All remaining
parameters are marginalized over. The combination of CMB, power spectrum
and weak lensing is already a powerful tool and a better knowledge
of one parameter doesn't improve much the constraints on $\beta^{2}$.
CMB alone, instead, improves by a factor 3 when $\Omega_{c}$ is known
and by a factor 2 when $h$ is known. The power spectrum is mostly
influenced by $\Omega_{c}$, which allows to improve constraints on
the coupling by more than a factor 2. Weak lensing gains the most
by a better knowledge of $\sigma_{8}$.

The constraint on $\beta^{2}$ can be easily converted into a constraint
on the Brans-Dicke coupling parameter $\omega_{BD}$ since (see e.g.
\cite{Pettorino:2008ez})
\begin{equation}
3+2\omega_{BD}=\frac{1}{2\beta^{2}}
\end{equation}
We obtain then
\begin{equation}
\omega_{BD}>800
\end{equation}
 The post-Newtonian parameter $\gamma_{PPN}$ is related (in the limit
$\gamma\ll1$) to $\omega_{BD}$ as $\gamma_{PPN}=1-2/(2\omega_{BD}+3)$
so that our final predicted constraint on $\beta^{2}$ reflects into
a constraint 
\begin{equation}
|\gamma_{PPN}-1|<1.2\times10^{-3}
\end{equation}
This is still two orders of magnitude weaker than the limits on a
scalar coupling to baryons obtained in laboratory or solar system
experiments, of the order of $|\gamma_{PPN}-1|<10^{-5}$ (see e.g.
\cite{2003Natur.425..374B,2007PhRvL..98b1101K,Nakamura:2010zzi})
but of course the cosmological bounds are complementary to the local
measurements since here we deal with the coupling to dark matter.

\begin{table*}
\begin{tabular}{llllll}
\begin{minipage}[c]{150pt}%
\flushleft Set of fixed parameters %
\end{minipage} & %
\begin{minipage}[c]{80pt}%
\flushleft CMB%
\end{minipage} & %
\begin{minipage}[c]{80pt}%
\flushleft $P(k)$ %
\end{minipage} & %
\begin{minipage}[c]{80pt}%
\flushleft WL %
\end{minipage} & %
\begin{minipage}[c]{100pt}%
\flushleft CMB + $P(k)$ + WL %
\end{minipage} & \tabularnewline
\hline 
(Marginalized on all params)  & 0.0094  & 0.0015  & 0.012  & 0.00032  & \tabularnewline
$P_{1}=\{\alpha\}$  & 0.0093  & 0.00085  & 0.0098  & 0.00030  & \tabularnewline
$P_{2}=\{P_{1}\}\cup\{\Omega_{c}\}$  & 0.0026  & 0.00065  & 0.0084  & 0.00030  & \tabularnewline
$P_{3}=\{P_{2}\}\cup\{h\}$  & 0.00054  & 0.00040  & 0.0076  & 0.00026  & \tabularnewline
$P_{4}=\{P_{3}\}\cup\{\Omega_{b}\}$  & 0.00033  & 0.00028  & 0.0037  & 0.00020  & \tabularnewline
$P_{5}=\{P_{4}\}\cup\{n_{s}\}$  & 0.00033  & 0.00024  & 0.0034  & 0.00019  & \tabularnewline
$P_{6}=\{P_{5}\}\cup\{\sigma_{8}\}$  & 0.00033  & 0.00024  & 0.0017  & 0.00019  & \tabularnewline
$P_{7}=\{P_{6}\}\cup\{\log(A)\}$  & 0.00032  & 0.00024  & 0.0017  & 0.00019  & \tabularnewline
\hline 
\end{tabular}\label{tab:fixedparlist} \caption{1-$\sigma$ errors for $\beta^{2}$, for CMB, $P(k)$, WL and CMB+$P(k)$+WL.
For each line the set of parameters $P_{i}$ has been fixed to the
reference value. The first line corresponds to the case in which we
have marginalized over all parameters.}
\end{table*}

\begin{table*}
\begin{tabular}{llllll}
\begin{minipage}[c]{150pt}%
\flushleft Fixed parameter %
\end{minipage} & %
\begin{minipage}[c]{80pt}%
\flushleft CMB%
\end{minipage} & %
\begin{minipage}[c]{80pt}%
\flushleft $P(k)$ %
\end{minipage} & %
\begin{minipage}[c]{80pt}%
\flushleft WL %
\end{minipage} & %
\begin{minipage}[c]{100pt}%
\flushleft CMB + $P(k)$ + WL %
\end{minipage} & \tabularnewline
\hline 
(Marginalized on all params)  & 0.0094  & 0.0015  & 0.012  & 0.00032  & \tabularnewline
$\alpha$  & 0.0093  & 0.00085  & 0.0098  & 0.00030  & \tabularnewline
$\Omega_{c}$  & 0.0026  & 0.00066  & 0.0093  & 0.00032  & \tabularnewline
$h$  & 0.0044  & 0.0013  & 0.011  & 0.00032  & \tabularnewline
$\Omega_{b}$  & 0.0087  & 0.0014  & 0.012  & 0.00030  & \tabularnewline
$n_{s}$  & 0.0074  & 0.0014  & 0.012  & 0.00028  & \tabularnewline
$\sigma_{8}$  & 0.0094  & 0.00084  & 0.0053  & 0.00030  & \tabularnewline
$\log(A)$  & 0.0090  & 0.0015  & 0.012  & 0.00032  & \tabularnewline
\hline 
\end{tabular}\caption{1-$\sigma$ errors for $\beta^{2}$, for CMB, $P(k)$, WL and CMB+$P(k)$+WL.
For each line, only the parameter in the left column has been fixed
to the reference value. The first line corresponds to the case in
which we have marginalized over all parameters.}
\label{tab:fixedpar} 
\end{table*}

\section{Conclusions}

\label{conclusions}

Future cosmological surveys will measure the properties of our universe
to a unprecedented precision. This will allow not only to measure
the main cosmological parameters but also to test models that challenge
the standard scenario of particle physics and cosmology. The two main
tools of observational cosmology, namely cosmic microwave background
and large-scale structure, can be combined to extract the maximal
amount of information.

In this paper we combined forecasts from the Planck CMB experiment
and from future redshift surveys and weak lensing surveys based on
the proposed Euclid satellite to get constraint on a model of coupled
dark energy. In this model, dark energy mediates a force acting on
dark matter and produces an additional attractive interaction which
modifies Einstein's gravity. The coupling parameter $\beta^{2}$ measures
the amount of this modification. We find that combining Planck with
a Euclid-like survey can constrain $\beta^{2}$ to a level of $3\cdot10^{-4}$,
two orders of magnitude better than current constraints \cite{amendola_quercellini_2003,LaVacca:2009yp};
this bound is weaker but complementary to the small-scale limits set
by local gravity tests on Yukawa corrections (see e.g. \cite{Nakamura:2010zzi,2007PhRvL..98b1101K}).
A better knowledge of $\Omega_{c}$, $h$, $n_{s}$ can improve the
precision of measurement of the coupling as obtained by CMB measurements,
due to manifest degeneracies among these parameters and the fifth
force as measured by $\beta^{2}$. Interestingly though, the combination
of CMB, power spectrum and weak lensing is already so powerful that
no much gain is obtained, overall, by a better knowledge of the other
parameters. \textbf{ } In this sense, these three probes are optimally
complementary probes of the dark energy-dark matter coupling.
\begin{acknowledgments}
Support was given to V.P. and C.Q. by the Italian Space Agency through
the ASI contracts Euclid-IC (I/031/10/0). V.P. is also supported by
the Young Scientist SISSA Grant. Support was given to L.A. by the
Deutsche Forschungsgemeinschaft through the programme TRR33 ``The
Dark Universe''. We acknowledge the Euclid Consortium for the preparation
of the Euclid Red Book, the Euclid Science Coordinators for reviewing
the draft of this paper and the Euclid Theory Working Group for very
fruitful exchange of ideas. We thank C. Carbone and C. Baccigalupi
for useful comments and discussion. 
\end{acknowledgments}
\bibliography{Bibdraft}

\end{document}